\documentstyle[preprint,aps]{revtex}
\begin{document}
\draft
\title
{\bf Is there a Pronounced Giant Dipole Resonance in $^4$He?}
\author{Victor D. Efros$^{1,2)}$, Winfried Leidemann$^{3)}$, and
Giuseppina Orlandini$^{3,4)}$}
\address{
1) European Centre for Theoretical Nuclear Physics and Related Areas,
Villa Tambosi, I-38050 Villazzano (Trento), Italy\\
2) Russian Research Centre "Kurchatov Institute", Kurchatov Square 1,
123182 Moscow, Russia\\
3) Dipartimento di Fisica, Universit\`a di Trento, 
 I-38050 Povo (Trento), Italy\\
4) Istituto Nazionale di Fisica Nucleare, Gruppo collegato di Trento, Italy
}

\date{\today}
\maketitle

\begin{abstract}
A four--nucleon calculation of the total $^4$He photodisintegration cross 
section is performed. The full final--state interaction is taken into account
for the first time. This is achieved via the method of the 
Lorentz integral transform. Semi--realistic NN interactions are employed. 
Different from the known partial two--body $^4$He$(\gamma,n)^3$He and 
$^4$He$(\gamma,p)^3$H cross sections our total cross section exhibits
a pronounced giant resonance. Thus, in contrast to older $(\gamma,np)$ data, 
we predict quite a strong contribution of the $(\gamma,np)$ 
channel at the giant resonance peak energy. 
\end{abstract}


The photodisintegration of $^4$He has received much attention
in the last 25 years.  Experimental work concentrated mainly on the two dominant
two--body breakup channels ($^3$He$ + n$, $^3$H$ + p$). In a first round 
of experiments a rather strong peak of the giant dipole resonance was found,
while more recent experiments find a much less pronounced peak.
The suppression of the two--body breakup peak was confirmed in 
four--nucleon calculations that take into account the
important final state interaction (FSI) via a semi--realistic NN potential
\cite{Hof92,Sand96}.
Much less is known about the total $\alpha$ photoabsorption cross section
($^4$He$ +\gamma \rightarrow X)$. In the vicinity and beyond the peak
there are neither theoretical calculations that take into account FSI
nor experimental total cross section measurements.

The situation for the $^4$He photodisintegration 
seems to be settled only for the two--body breakup channels at lower
energies. Yet the results are rather puzzling because it is not
understood why the $\alpha$-particle should have such a suppressed
giant dipole resonance.
Cross sections for transitions to other channels $(\gamma,pn)d$, 
$(\gamma,2p2n)$ and $(\gamma,d)d$ obtained in the older experiment 
\cite{Gorb76,Ark78} are very small and cannot influence the general picture
at all.
Furthermore, the new $(\gamma,p)^3$H and $(\gamma,n)^3$He
data combined with those cross sections would lead to a bremsstrahlung weighted 
sum over the photoabsorption spectrum which is substantially lower than the 
well known model--independent sum rule estimate. A theoretical calculation 
of the total photoabsorption
cross section would certainly help to get a better understanding of
these problems, since the giant resonance is in principle a
feature of the total cross section.

In the present work the theoretical calculation of the total
cross section  is carried out with consideration of the full FSI. Previously 
the FSI was taken into account completely only below 
the three--body $p+n+d$ breakup threshold $E_{\gamma}=26.1$ MeV 
\cite{Hof92,Sand96}. For the two--body breakup the resonating group 
calculation of Ref. \cite{Hof92} was extended to somewhat higher energies 
taking  into account FSI due to other channels approximately. At  
$E_{\gamma}>50$ MeV the two--body reactions were treated in the  
plane--wave approximation \cite{Sand96}. 

We calculate the total photoabsorption cross section in the whole energy range 
below the pion threshold. We consider the E1 transition in the long--wavelength
limit using the unretarded dipole transition operator 
\[ \vec D = \sum_{i=1}^Z (\vec r_i - \vec R_{cm})\,. \] 
In this way we take into account meson exchange currents via the Siegert 
theorem. The E2 contributions to the total cross section are small even at 
high photon energy \cite{Sand96} and they tend to cancel with the E1 
retardation contributions \cite{Ger64}. Our nuclear hamiltonian includes 
central even local NN potentials and the Coulomb interaction.

We can write down the total photoabsorption cross section as
\[ \sigma_{tot}(E_{\gamma})=4\pi^2(e^2/\hbar c)E_{\gamma}R(E_{\gamma})\,, \]
where $R$ is the dipole response function,
\[ R(E_{\gamma})=\int df |\langle\Psi_f|D_z|\Psi_0\rangle|^2\delta(E_f-E_0-
E_{\gamma})\,. \] Here $\Psi_0$ is the $\alpha$-particle wave function and 
$\Psi_f$ are final state wave functions normalized as 
$\langle\Psi_f|\Psi_{f'}\rangle=\delta(f-f')$.  In the above relations we 
neglect the very small nuclear recoil energy. We calculate the response 
function $R$ via evaluation and subsequent inversion of its Lorentz integral 
transform, a method we proposed for the response of an arbitrary $N$ particle
system to an external probe \cite{ELO94}. The method has already been 
successfully applied for obtaining the accurate longitudinal $(e,e')$ 
response functions of the two--, three--, and four--nucleon systems 
\cite{ELO94,Sara95,ELO96}. The transform $\cal{L}(\sigma)$ 
of the response $R$ is found as
\begin{equation} 
{\cal L}(\sigma)=\langle \tilde{\Psi}(\sigma)| \tilde{\Psi}(\sigma)\rangle, 
\end{equation}
$\tilde{\Psi}$ being the solution to the Schr\"odinger--like
equation 
\begin{equation}
(\hat{H}-E_0+\sigma)\tilde{\Psi}(\sigma)=Q
\end{equation}
with the source--term $Q=D_z\Psi_0$. The function $\tilde{\Psi}$ is localized
and continuum calculations are thus avoided in our approach.

We use the same NN potential model, Trento (TN) potential, as in our work 
on the longitudinal response function \cite{ELO96}. We also consider the 
Malfliet--Tjon (MT) I+III potential \cite{MT69}, which was used in Ref. 
\cite{Sand96} for calculating the reaction $\gamma + ^4$He $\rightarrow 
^3$He $+ n$. We use the value of $\lambda=1.555$ fm$^{-1}$ entering the 
attractive part of the MT potential as listed in Ref. \cite{KG92}. This value 
just leads to correct low--energy parameters of NN scattering as given in 
Ref. \cite{MT69}. In some $^4$He bound--state calculations the value 
$\lambda=1.55$ fm$^{-1}$  listed in Ref. \cite{MT69} has been used that 
leads to an increase in the $E_b(^4$He) value by about 1.4 MeV.

Our $\alpha$-particle wave function $\Psi_0$ is an
eigensolution for the same NN potential. The corresponding matter r.m.s. radia
and binding energies are 1.41 fm and 30.5 MeV for the TN 
potential, and 1.43 fm and 29.2 MeV for the MT potential. The latter value
is close to those reported in the literature, see Ref. \cite{Sand96}. 
The binding energies are reasonable as compared to the experimental value of 
28.3 MeV, and the radia are close to the experimental value of 1.45 fm.

In Fig. 1 we show the $s$-wave phase shifts (no Coulomb interaction
included for $^1S_0$) of both potential models in comparison to those 
of a realistic interaction (Paris 
potential \cite{Lac80}). It is evident that MT and TN potentials
do not lead to significantly different phase shifts than the Paris
potential.
The $^1S_0$ scattering 
length  equals to --17.9 fm for the TN potential and --23.3 fm for 
the MT potential, so the MT potential is a little bit more attractive in the 
$^1S_0$ channel than the TN potential. The TN scattering length is close to the 
value of nn and pp (no Coulomb force) scattering ($a_{nn}(^1S_0)=-17.6$ fm 
for Paris potential), while the MT scattering 
length is close to that of np scattering ($a_{np}(^1S_0)=-23.7$ fm).

We solve Eq. (2) for $L=T=1$ and $S=0$ with 
the help of the correlated hyperspherical expansion and the hyperradial 
expansion of the same form as in Ref. \cite{ELO96}. The $K_{max}$ value 
equals to 7. The $\sigma$ value in Eqs. (1), (2) is of the form 
$-\sigma_R+i\sigma_I$ with $\sigma_I=const$, and the values of $\sigma_I=20$ 
MeV and 5 MeV have been employed. In Fig. 2 the convergence of the transform, 
Eq. (1), with respect to $K_{max}$ is shown for $\sigma_I=20$ MeV for the MT 
potential. While inverting the transform the true low energy behavior 
$[E_{\gamma}-(E_{\gamma})_{min}]^{3/2}$ have been incorporated into our trial 
response. The inversion has been performed both for $\sigma_I=20$ MeV and for 
a combination of the transforms with $\sigma_I=5$ and 20 MeV chosen so that the 
former transform gives a predominant contribution to the very steeply rising 
low energy wing of the response and the latter ones to its high energy wing. 
The responses obtained in these two versions practically coincide with each 
other. The transforms in Fig. 2 with $K_{max}=5$ and 7 lead to practically 
identical responses and that for $K_{max}=3$ is also not very different. For 
the TN potential one finds a similarly good convergence in $K_{max}$ as well. 

Besides the checks of the convergence, the overall test of the final results 
is provided by sum rule calculations. We compare the bremsstrahlung weighted 
sum $\sigma_b=\int_{E_{\gamma}^{th}}^{\infty} \sigma_{tot}(E_{\gamma})
E_{\gamma}^{-1}dE_{\gamma}$ and the TRK sum $\sigma_{TRK}=
\int_{E_{\gamma}^{th}}^{\infty} \sigma_{tot}(E_{\gamma})dE_{\gamma} = 
59.74(1+\kappa)$ MeV mb calculated with our cross sections with an independent 
calculation of these quantities using the sum rules ( $\sigma_b = 
4\pi^2(e^2/\hbar c)\langle \Psi_0 | D_z D_z |
\Psi_0\rangle$, $\kappa = \langle \Psi_0 |
[D_z,[V,D_z]] | \Psi_0 \rangle (m/\hbar^2)A/NZ$).
The sum rule values are $\sigma_b=2.41$ mb, $\kappa=0.727$ for the TN 
potential and $\sigma_b=2.48$ mb, $\kappa=0.684$  for the MT potential. 
By integrating our cross sections explicitly we obtain $\sigma_b=2.40$ mb, 
$\kappa=0.754$ for the TN potential and $\sigma_b=2.48$ mb, $\kappa=0.712$ 
for the MT potential. 
The agreement of the $\sigma_b$ values with the sum rules is perfect that 
reflects a good accuracy of the low energy wings of the responses obtained. 
The resulting relative deviations from the TRK sum rule are about
1.5\% for both potentials.

One may note that the $\kappa$ values for the potentials we use are lower 
than those provided by fully realistic  NN interactions. The latter values 
range from 1.0 to 1.3 \cite{HAH78,Gari78,Sch87}, thus we underestimate 
$\sigma_{TRK}$ by 15-25\%. We believe that the main part of the missing 
strength should lead to an increase of the cross section at higher energies,
while our potential models should provide quite realistic results up to the 
pion threshold. In fact a rough estimate of $\sigma_b$, which we performed 
for realistic NN interactions, is close to the $\sigma_b$ values for our 
potentials. In any case, an increase in the $\kappa$ value would only 
strengthen our conclusions about the strong $(\gamma,np)$ cross section
which we predict below.
 
At this point we should mention that
our calculation is performed consistently with our semi--realistic 
hamiltonians, i.e. applying the Siegert theorem we use the energy
eigenvalues of the hamiltonian.  However for comparison with
experiment we perform the shift $\sigma_{tot}(E_{\gamma})
\rightarrow \sigma_{tot}(E_{\gamma}+\Delta E_b)$, $\Delta E_b$ being the 
difference of the calculated and experimental binding energies. In this way
we obtain the proper breakup threshold thus correcting for some overbinding 
of our $\alpha$-particle.

Unfortunately there are no direct experimental data on the $^4$He total 
photoabsorption cross section. Nevertheless we would like to make
a comparison with experimental data. Therefore we proceed
as follows. For the low--energy region we make interpolations of the 
$(\gamma,n)$ data from \cite{Ber80} and the $(\gamma,p)$ data from 
\cite{Feld90} and sum up the resulting $(\gamma,p)$ and $(\gamma,n)$ cross 
sections (dotted curves in Figs. 3,4).
Since the $(\gamma,d)d$ cross section can be safely neglected
(see e.g. \cite{Ed95}) this should lead to a rather good estimate for the
total cross section below the three--body breakup threshold. Furthermore, we 
also show the cross sections of other low--energy experiments 
\cite{Ward81,Bern88,Asai94}. Assuming that $(\gamma,p)$ and $(\gamma,n)$ cross 
sections are more or less equal we double the experimental cross sections in 
order to have further estimates for the two--body breakup.
Beyond 26.1 MeV they represent lower experimental bounds for the total
cross section. In Fig. 3 these estimates are shown together with the 
calculated cross sections for MT and TN potentials. There is a rather good 
agreement of our responses with the estimated  
experimental two--body cross section up to the three--body breakup threshold. 
The MT potential leads to a slightly higher low--energy cross section than the 
TN potential that may be related to the somewhat stronger attraction in the NN 
$^1S_0$ channel. For the MT potential we find a similar agreement with 
experimental data as was found in \cite{Sand96} for the same potential 
model for the $(\gamma,n)^3$He channel.

Beyond the three--body breakup threshold our  cross sections reveal further 
increase. Since theoretical as well as experimental results for the 
$(\gamma,p)$
and $(\gamma,n)$ cross sections show a flattening beyond the three--body 
threshold, the further increase has to be attributed to $(\gamma,np)$ 
reactions. Thus the $(\gamma,np)$ channel increases the peak of the giant 
dipole resonance considerably. As can be seen in Fig. 4 it
leads to a rather pronounced resonance peak.
 Also in the high--energy sector we show 
lower experimental bounds for the total cross section. They consist of the
sum of the $(\gamma,p)^3$H and $(\gamma,n)^3$He cross sections from 
Ref. \cite{Gorb76}
and the doubled $(\gamma,p)^3$H data from Ref. \cite{Jones91}.
From the comparison of these estimates with our theoretical total cross 
sections one would expect quite an important contribution of the 
$(\gamma,np)$ channel in the whole energy range.

Finally, we summarize our work. For the first time the total cross section 
of the $\alpha$-particle photodisintegration was calculated in the framework
of four--nucleon dynamics with full FSI. The results show a very pronounced 
peak of the giant dipole resonance. Therefore it seems that a typical many-body
feature emerges also from a genuine few-body calculation of the four-nucleon
system. The peak is considerably higher than the 
sum of the cross sections of the two important two--body breakup channels 
($^3$H+$p$, $^3$He+$n$). Thus we predict quite a strong contribution
of the $(\gamma,np)$ channel already at rather low energies. More 
experimental work is needed to confirm this prediction. At 
present some data on $(\gamma,np)$ with high statistics are available only 
beyond 80 MeV \cite{Doran93}, while the energy range between three--body 
breakup threshold and 80 MeV remains to be explored.

The authors thank H.M. Hofmann for helpful correspondence.

\vfill\eject

\begin{figure}
\caption{NN scattering phase shifts of the partial waves $^1S_0$ (a) 
and $^3S_1$ (b) for the following potentials: TN (dashed curves), 
MT (dotted curves), and Paris (full curves).}
\end{figure}

\begin{figure}
\caption{
The Lorentz transform for the MT potential with various $K_{max}$ values.}
\end{figure}

\begin{figure}
\caption{
Theoretical results for the total $^4$He photoabsorption cross section 
at low energy with MT (dashed curve) and TN potentials (full curve).
Also shown is the estimate for the two--body breakup (dotted curve with 
typical size of the experimental error), which is based on the experimental 
results of Refs. [15,16] as well as doubled experimental cross 
sections for $(\gamma,p)$ [19] (open circles) and for $(\gamma,n)$ 
[18] (triangles) and [20] (full circles) (for further 
explanation see text). The three-body breakup threshold is marked by an arrow.}
\end{figure}

\begin{figure}
\caption{
As Fig. 3, but for an extended energy range up to 140 MeV.
Estimate for lower experimental bound (dotted curve) and additional lower
bound estimates with data from [3] (diamonds), [19] 
(open circles), and [21] (squares) (for further explanation see text).}
\end{figure}


\begin{thebibliography}{99}
\bibitem{Hof92} M. Unkelbach and H. M. Hofmann, Nucl. Phys. {\bf A549}, 550
(1992).
\bibitem{Sand96} G. Elkermann, W. Sandhas, S. A. Sofianos, and H. Fideldey, 
Phys. Rev. C {\bf 53}, 2638 (1996), and references therein.
\bibitem{Gorb76} A. N. Gorbunov, Proceedings of the P.N. Lebedev Physics
Institute, {\bf 71}, 1 (1976). 
\bibitem{Ark78} Yu. M. Arkatov, P. I. Vatset, V. I. Voloschuk, V. N. Gur'ev, 
and A. F. Khodjachih, Ukr. Fiz. Zh. {\bf 23}, 1818 (1978) (in Russian) [Ukr. 
Phys. J.].
\bibitem{Ger64} S. B. Gerasimov, Phys. Lett. {\bf 13}, 240 (1964).
\bibitem{ELO94} V. D. Efros, W. Leidemann, and G. Orlandini, Phys. Lett. 
{\bf B 338}, 130 (1994).
\bibitem{Sara95} S. Martinelli, H. Kamada, G. Orlandini, and W. Gl\"ockle, 
Phys. Rev. C {\bf 52}, 1778 (1995).
\bibitem{ELO96} V.D. Efros, W. Leidemann, and G. Orlandini, Phys. Rev. Lett.
{\bf 78}, 432 (1997).
\bibitem{MT69} R. A. Malfliet and J. Tjon, Nucl. Phys. {\bf A127}, 161 (1969).
\bibitem{KG92} H. Kamada and W. Gl\"ockle, Nucl. Phys. {\bf A548}, 205 (1992).
\bibitem{Lac80} M. Lacombe, B. Loiseau, J. M. Richard, R. Vinh Mau, J. C\^ot\'e,
P. Pir\`es, and R. de Tourreil, Phys. Rev. {\bf C21}, 861 (1980).
\bibitem{HAH78} W. Heinze, H. Arenh\"ovel, and G. Horlacher, Phys. Lett.
{\bf B76}, 379 (1978). 
\bibitem{Gari78} M. Gari, H. Hebach, B. Sommer, and J. G. Zabolitzky,
Phys. Rev. Lett. {\bf 41}, 1288 (1978).
\bibitem{Sch87} R. Schiavilla, A. Fabrocini, and V. R. Pandharipande,
Nucl. Phys {\bf A473}, 290 (1987).
\bibitem{Ber80} B. L. Berman, D. D. Faul, P. Meyer, and D. L. Olson,
Phys. Rev. {\bf C22}, 2273 (1980).
\bibitem{Feld90} G. Feldman, M. J. Balbes, L. H. Kramer, J. Z. Williams,
H. R. Weller, and D. R. Tilly, Phys. Rev. {\bf C42}, 1167 (1990).
\bibitem{Ed95} E. L. Tomusiak, W. Leidemann, and H. M. Hofmann, Phys. Rev.
{\bf C52}, 1963 (1995).
\bibitem{Ward81} L. Ward, D. R. Tilly, D. M. Skopik, N. R. Robertson and
H. R. Weller, Phys. Rev. {\bf C24}, 317 (1981).
\bibitem{Bern88} R. Bernabei et al., Phys. Rev. {\bf C38}, 1990 (1988).
\bibitem{Asai94} J. Asai et al., Few-Body Syst. Suppl. {\bf 7}, 136 (1994).
\bibitem{Jones91} R. T. Jones, D. A. Jenkins, P. T. Debevec, P. D. Harty,
and J. E. Knott, Phys. Rev. {\bf C43}, 2052 (1991).
\bibitem{Doran93} S. M. Doran et al., Nucl. Phys. {\bf A559}, 347 (1993).
\end{thebibliography}
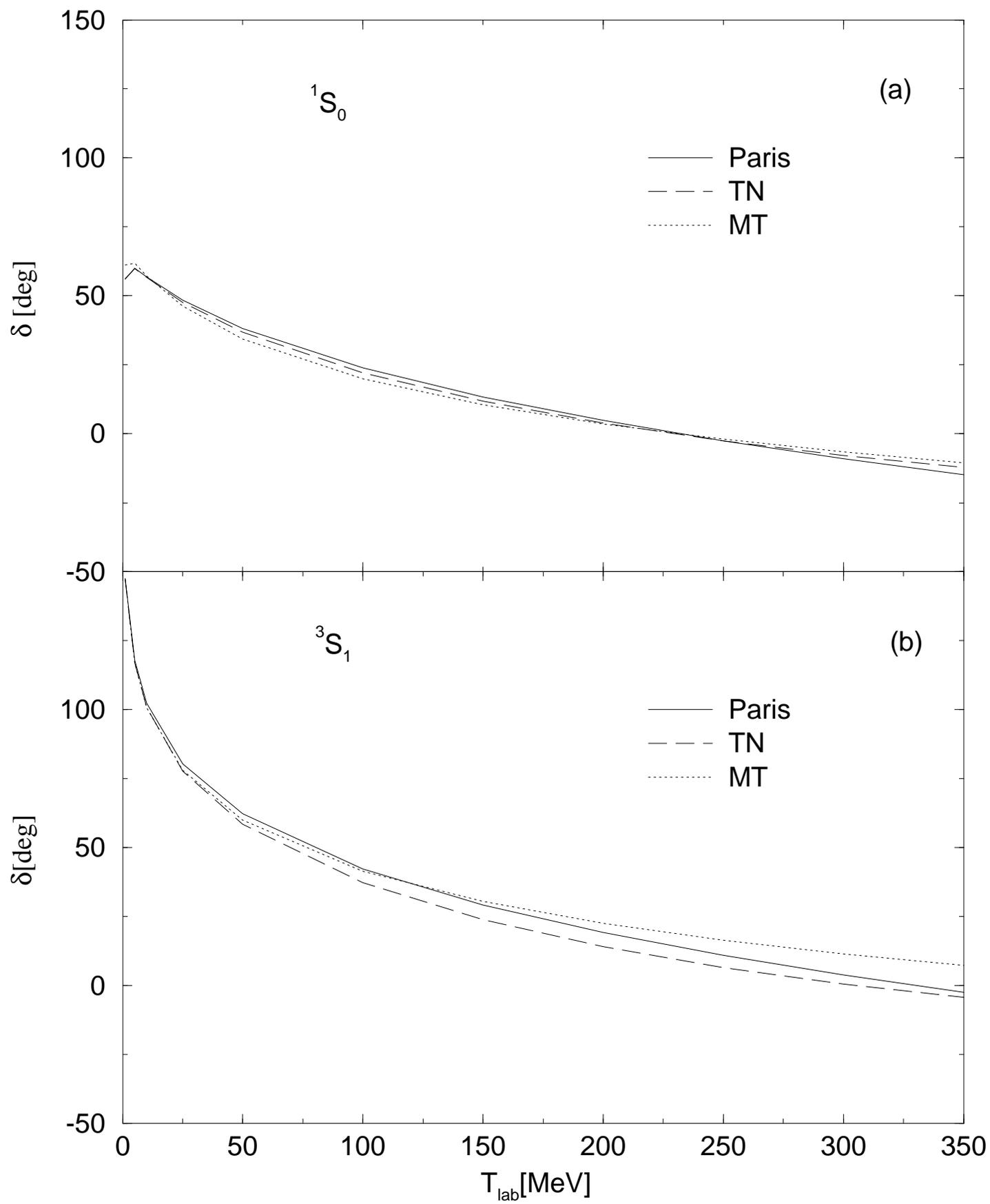
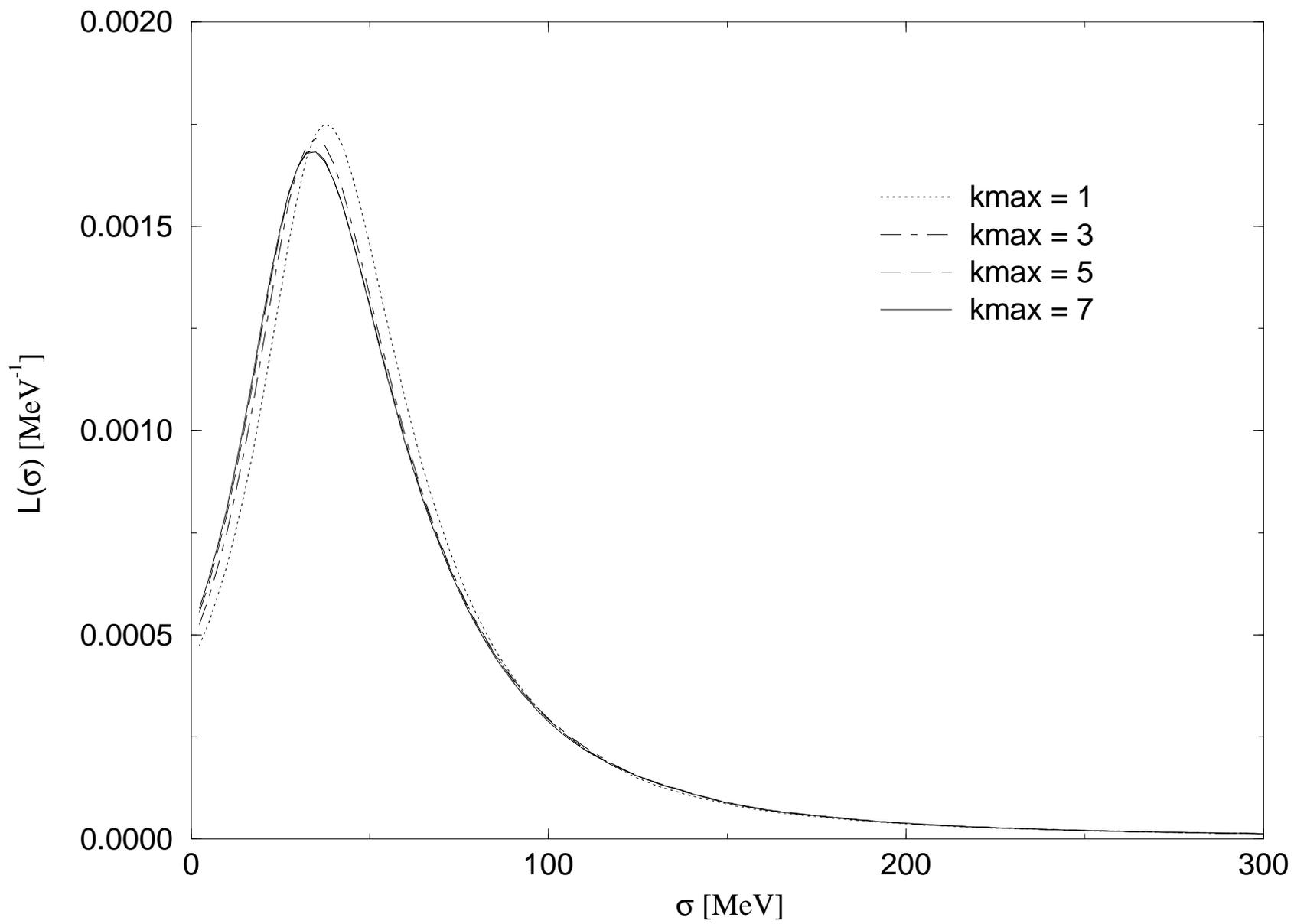
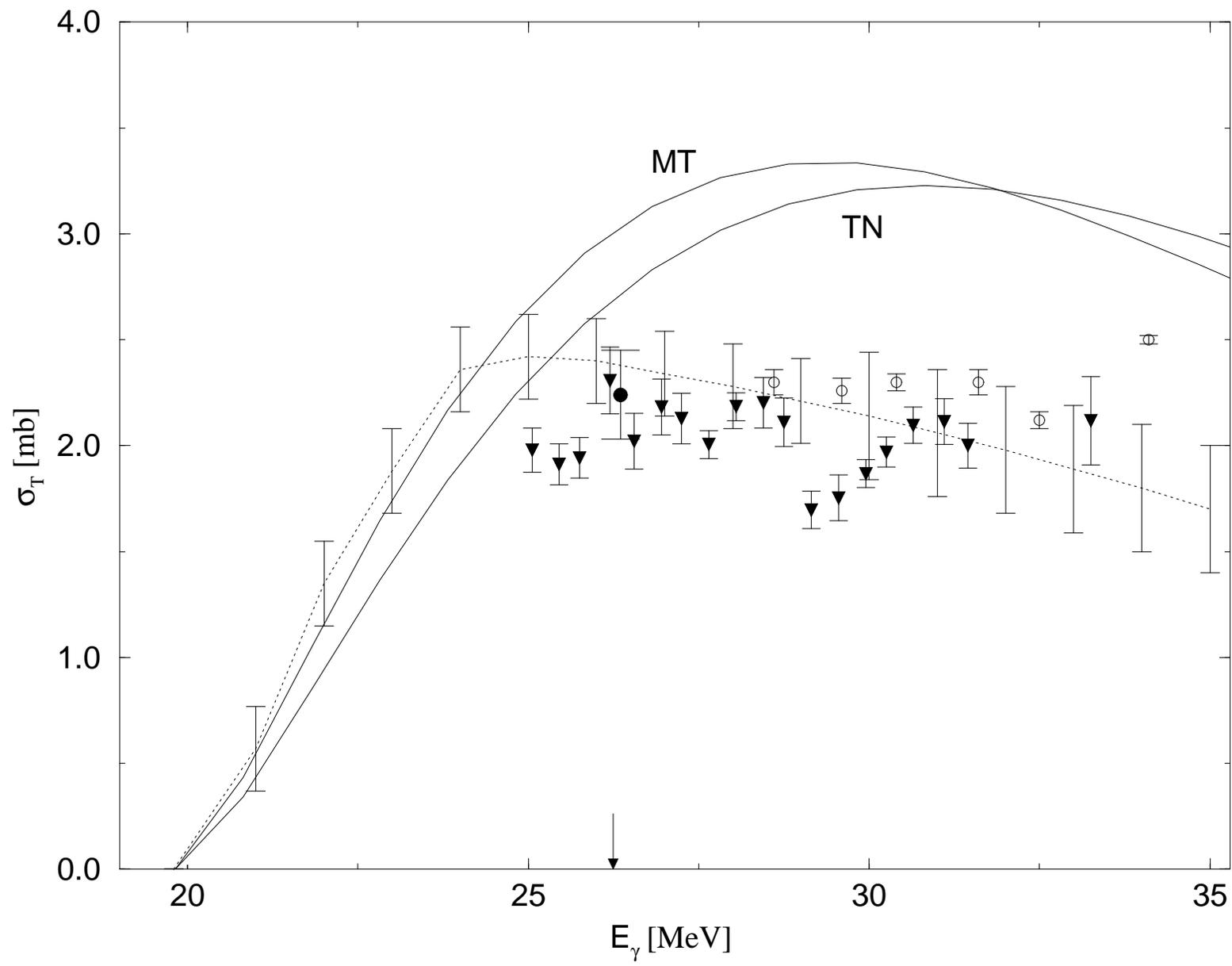
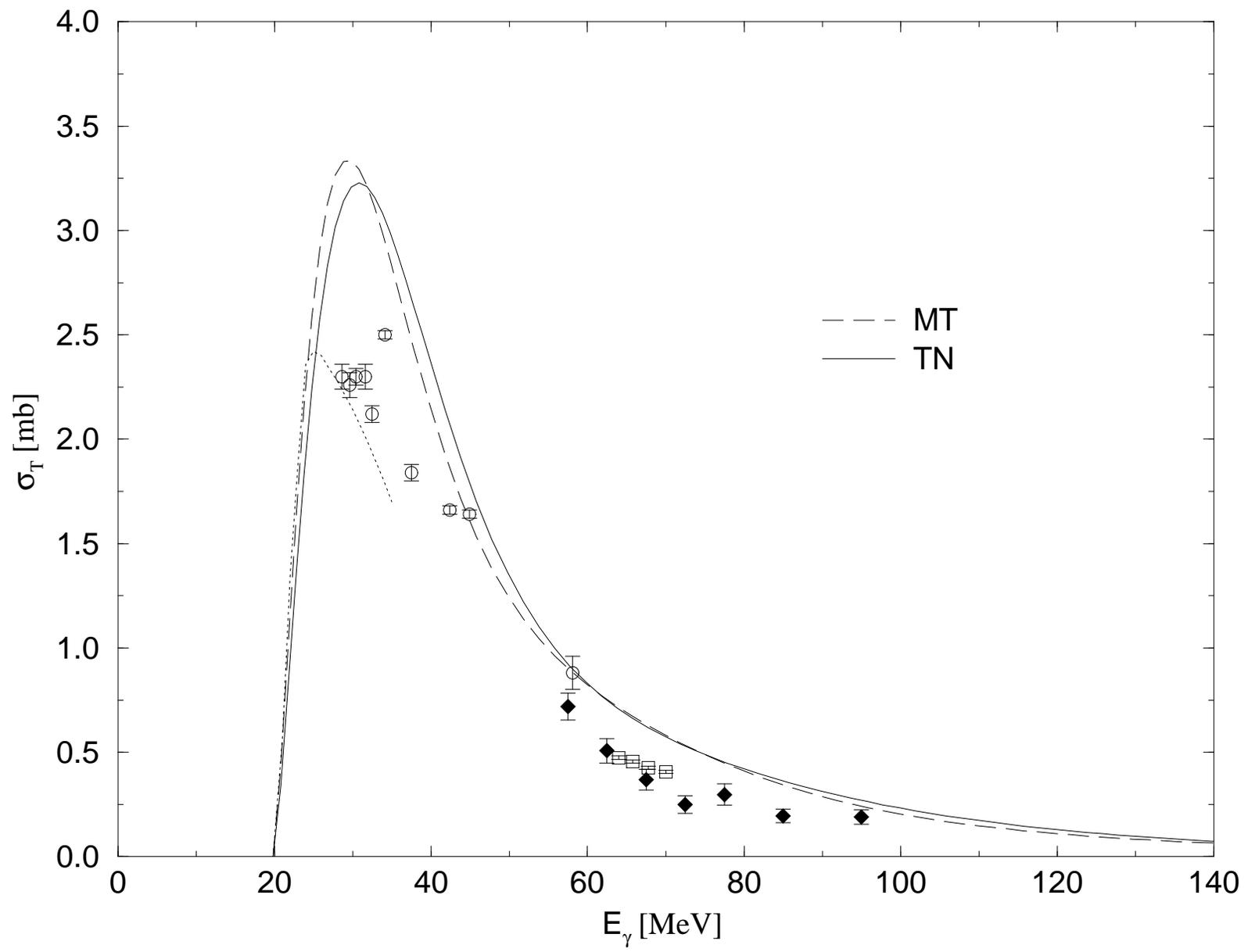
\end{document}